\documentclass{PoS}

\usepackage{graphicx}

\usepackage[utf8]{inputenc}
\usepackage[T1]{fontenc}

\usepackage{amsfonts}
\usepackage{amssymb}
\usepackage{tikz}

\usepackage{soul} 

\definecolor{col5}{RGB}{139,0,0}

\title{Testing dynamic stabilisation in complex Langevin simulations}

\ShortTitle{Testing dynamic stabilisation in complex Langevin simulations}

\author{Felipe Attanasio$^{1}$, \speaker{Benjamin
Jäger}$^{1,2}$
\\
$^{1}$Department of Physics, College of Science, Swansea University, United Kindgom\\
$^{2}$ETH Zürich, Institute for Theoretical Physics, Wolfgang-Pauli-Str. 27,
8093 Zürich, Switzerland\\
E-mail:
\email{bejaeger@phys.ethz.ch} 
}

\abstract{ 
Complex Langevin methods have been successfully applied in theories that suffer
from a sign problem such as QCD with a chemical potential. We present and
illustrate a novel method (dynamic stabilisation) that ensures that Complex
Langevin simulations stay close to the SU(3) manifold, which lead to correct and improved results in
the framework of pure Yang-Mills simulations and QCD in the limit of heavy
quarks.}

\FullConference{34th annual International Symposium on Lattice Field Theory\\
24-30 July 2016\\
University of Southampton, UK}

\begin{document}
  
\section{Introduction}

The famous \emph{sign problem} is encountered in many areas of physics. One of
the most prominent examples is QCD with non-zero baryon chemical potential. The
fermion determinant leads to a complex weight in the associated Euclidean path
integrals, and thus prevents direct determination of the QCD phase diagram with
standard Monte Carlo techniques. Complex Langevin simulations
have been shown to enable simulations even when the sign problem is
severe~\cite{Aarts:2008rr, Sexty:2013ica,
Aarts:2014bwa,Sinclair:2015kva,Nagata:2016vkn}.
The hope is that fully dynamical QCD can be studied with this stochastic
quantization, which was first introduced in~\cite{Parisi:1980ys, Parisi:1984cs, Klauder:1983nn, Klauder:1983zm,
Klauder:1983sp}. A crucial step forward was made by employing 
{\em gauge cooling}~\cite{Seiler:2012wz,Aarts:2013uxa}, which is
used to control the distance from the unitary sub-manifold. Gauge cooling
made it possible to obtain the first results in QCD, both with 
heavy~\cite{Aarts:2015yba,Aarts:2016qrv} and with lighter
quarks~\cite{Sexty:2013ica, Aarts:2014bwa}. However, as already noted
in~\cite{Aarts:2016qrv}, gauge cooling is less effective at smaller values of the gauge 
coupling $\beta$, i.e.\ on coarser lattices. While this is not a problem per se, 
eventually the continuum limit will have to be taken, it limits the 
applicability to lower temperatures, which would require very large lattices.
Hence it is worthwhile to study modifications of the CL process, 
which ameliorate the process at smaller $\beta$ values, which will be presented
in the following.

\section{Dynamic stabilisation}

A generic first-order Langevin update step in Langevin time $\theta$ can be
written as~\cite{Damgaard:1987rr}
\begin{equation}
	U_{x,\nu}(\theta + \varepsilon) = \exp \Big[ \mathrm{i} \, \lambda^a \left( 
	- \varepsilon \, D^a_{x,\nu} \, S  + \sqrt{\varepsilon} \,
	\eta^a_{x,\nu} \right) \Big]\, U_{x,\nu}(\theta), 
\end{equation} 
where $\lambda^a$ are the Gell-Mann matrices and the action $S$ contains the
Yang-Mills gauge part and the logarithm of the fermion determinant as
\begin{equation}
S = S_{\rm YM} - \ln\,\det \,M.
\end{equation} 

For simplicity, we skip the discussion about the logarithm of
the fermion determinant and its potential problems on the convergence 
of the Langevin process. Detailed discussion can be found
in~\cite{Mollgaard:2013qra,Splittorff:2014zca,Nishimura:2015pba,Greensite:2014cxa}.
The drift of the action, i.e. $- D^a_{x,\nu} \, S$, governs the dynamics of this
stochastic process, which is realized by the Gaussian white
noise $\eta^a_{x,\nu}$. To circumvent the complexity of the Euclidean path
integral and evade the sign problem, we extend the gauge group from SU$(3)$ to SL$(3,\mathbb{C})$, 
so that all links can symbolically written as 
\begin{equation}
 U_{x,\nu} = \exp \Big[ \mathrm{i} \,a\, \lambda^{c} \left( A^{c}_{x,\nu} +
 \mathrm{i} \,B^{c}_{x,\nu}\right) \Big],
\end{equation} 
where $a$ is the lattice spacing and $A^{c}_{x,\nu}$ ($B^{c}_{x,\nu}$) are the
real (imaginary) coefficients of the Gell-Mann matrices. For SU$(3)$ gauge
links the imaginary components vanish completely, i.e.\ $B^{c}_{x,\nu} = 0$.  A
measure of the distance to SU$(3)$ sub-manifold is given by the unitarity
norm, defined as
\begin{equation}
d_2 = \frac{1}{3\,V} \sum_{x, \nu} \mathrm{Tr}\Big( U_{x,\nu} U_{x,\nu}^\dagger
-1 \Big)^2.
\end{equation} The main idea
of our new method, dynamic stabilisation, is to add a force to the Langevin
dynamics, i.e.
\begin{equation}
			U_{x,\nu}(\theta + \varepsilon) = \exp \Big[ i \lambda^a \left( 
	- \varepsilon \, D^a_{x,\nu} \, S + \textcolor{col5}{i \,\varepsilon\, \alpha_{DS}\,
	M^{a}_{x}} + \sqrt{\varepsilon} \, \eta^a_{x,\nu} \right) \Big]
	U_{x,\nu}(\theta),
\end{equation} that formally vanishes in the
continuum limit and provides a drift that is directed towards the SU$(3)$
sub-manifold. A possible solution is given by~\cite{Aarts:2016qhx}
\begin{equation}
M^a_x = i \, b^a_x \, \Big( \sum_c b^c_x\, b^c_x \Big)^3 \qquad b^a_x =
	\mathrm{Tr}
	\Big[ \lambda^a \sum_\nu U_{x,\nu} U^\dagger_{x,\nu} \Big],
\end{equation}
where the sum over $c$ is written explicitly for clarification. The additional
forces are independent of the direction of the link and hence are applied
equally in all four directions.  We point out that this solution is not unique
and many different choices can be found. We note here that the additional term
in the drift is not invariant under SL$(3,\mathbb{C})$ gauge transformations 
and not holomorphic, and hence cannot be derived from an action principle. However it is still invariant
under SU$(3)$ transformations. Instead it should be viewed as an additional
contribution to the drift, whose role is to contain the dynamics in the
non-compact direction. Due to the presence of this term the standard
justification of Complex Langevin~\cite{Aarts:2009uq,Aarts:2011ax} can no longer
be used, since the direct connection between the drift and complex weight is jeopardised. However,
an expansion in powers of the lattice spacing shows that
\begin{equation}
		M^{a}_x \sim a^7 \, \Big(\overline{B}^c_y\,\,
		\overline{B}^c_y\Big)^3 \, \overline{B}^a_x + \mathcal{O}(a^8)\,\quad
		\mathrm{ with }\,\quad \overline{B}^a_x = \sum_\mu B^a_{x,\mu},\,
\end{equation}
and hence the additional term is indeed formally irrelevant in the
continuum limit. We emphasise that this is very different from gauge
cooling, where the dynamics is controlled via SL$(3,\mathbb{C})$ gauge
transformations and not via a modification of the drift. In spite of this
open theoretical issue, we will see below that the results obtained with
DS are very promising. 
 
\begin{figure}[!ht]
	\centering
	\begin{minipage}{0.93\linewidth}
	\centering
	\includegraphics[width=\linewidth]{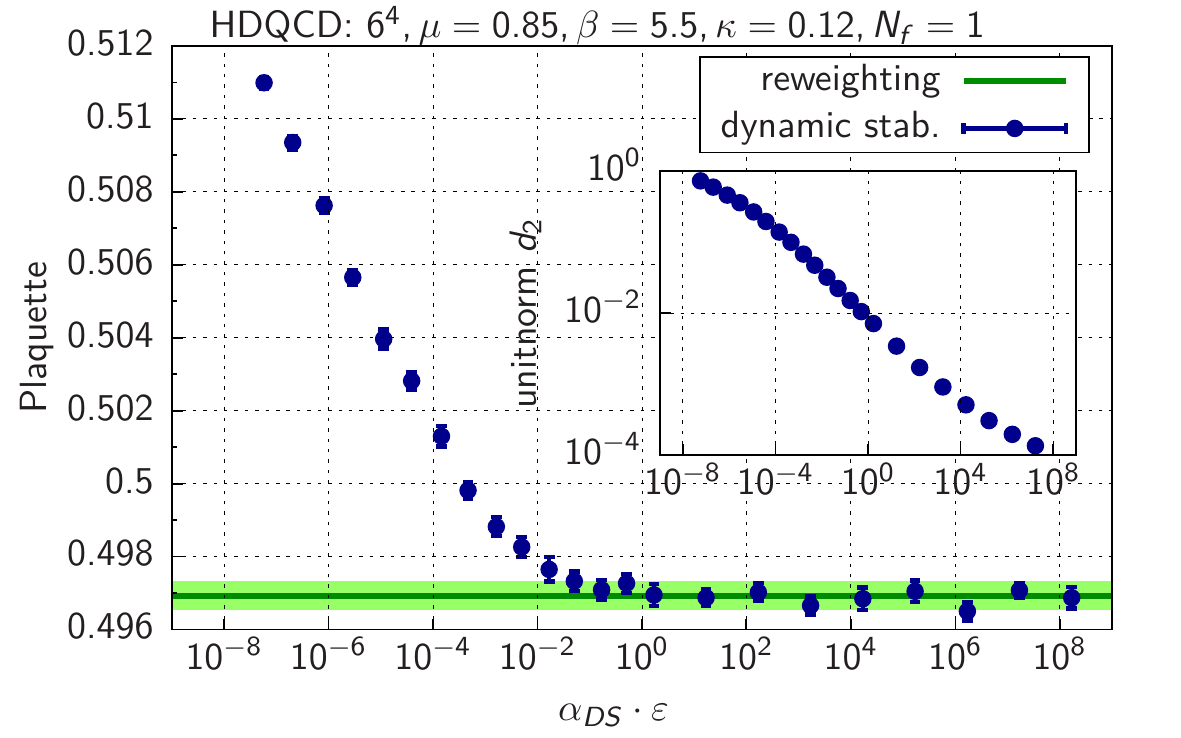}
	\end{minipage}
\caption{The plaquette as a function of the dynamic stabilisation coefficient
$\alpha_{DS}$ times the average stepsize $\varepsilon$. For sufficient
large values $\gtrsim 10^{0}$, we find perfect agreement with reweighting
results. The unitarity norm is shown in the small sub-section of this figure as
a function of the same $\alpha_{DS}\cdot \varepsilon$.}
\label{Fig1}   
\end{figure} 
The coefficient $\alpha_{DS}$ allows us to vary the
strength of the additional force in order to tune the distance to SU$(3)$. For very large values of
$\alpha_{DS}$ this relates to a dynamic re-unitarization of the gauge links. In the limit of small
coefficients the drift becomes irrelevant for the dynamics. The right choice of
the parameter $\alpha_{DS}$ depends on the application and needs some tuning. 
Dynamic stabilisation can be trivially combined with gauge cooling and we found
that one step of gauge cooling is very beneficial and sufficient for the
overall convergence. Therefore we use a single step of gauge cooling in all
simulation results shown in the following. Figure~\ref{Fig1} shows the plaquette
as a function of the dynamic stabilisation coefficient $\alpha_{DS}$ times the
average stepsize $\varepsilon$ for a HDQCD simulation.
For small values the results significantly deviate from reweighting results,
further illustrating that the unitarity norm is too large, which is shown in
the .
For $\alpha_{DS} \gtrsim 10^{0}$ the plaquette is in perfect agreement with
reweighting data. Please note that the force added to the drift scales with a
high powers of $\overline{B}^c_y$ and thereby indirectly with unitarity norm,
which is small for large combinations of $\alpha_{DS}\cdot \varepsilon$.

\begin{figure}[!ht]
	\centering
	\begin{minipage}{0.93\linewidth}
	\centering
	\includegraphics[width=\linewidth]{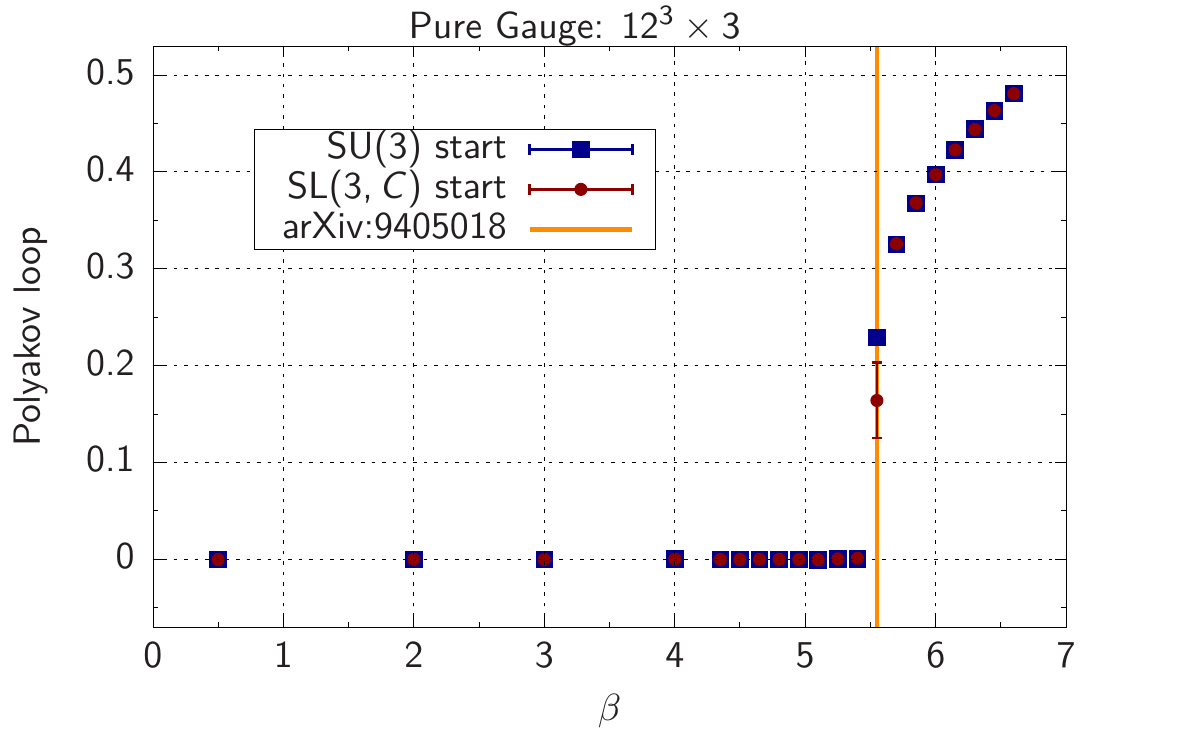}
	\end{minipage}
\caption{Pure Yang-Mills simulation using dynamic stabilisation as a function of the gauge
coupling $\beta$ for two different initial conditions. The result from Hybrid
Monte Carlo~\cite{Cella:1994sx} have been added for comparison.}
\label{Fig2} 
\end{figure}
For pure Yang-Mills simulations, the complexification of the gauge degrees of freedom
is not necessary, since the problem is free of a sign problem and
standard Monte Carlo techniques can be applied in a straightforward way. 
However, it serves as a good testing scenario to check that complex Langevin
simulations still work. By using SL$(3,\mathbb{C})$ gauge
links, we effectively use too many degrees of freedom and allow the system to generate imaginary drifts, 
which ideally would exactly vanish. However, round off errors 
and subsequent small imaginary components will generate non-zero $B^{c}_{x,\nu}$.
We previously showed that complex Langevin simulations using just gauge cooling
indeed reproduce the correct deconfinement transitions, when we start with an unitary
configuration, i.e\ $B^{c}_{x,\nu} = 0$ and use gauge cooling to maintain a
sufficient small unitarity norm ~\cite{Aarts:2015hnb}. However, using a generic 
SL$(3,\mathbb{C})$ start, with $B^{c}_{x,\nu} \neq 0$, leads to a rising
unitarity norm along the Langevin time until the simulations fail. Adding 
dynamic stabilisation correctly reproduce the deconfinement transition as
obtained in ~\cite{Cella:1994sx}. Figure~\ref{Fig2} shows such a comparison of
pure Yang-Mills simulation using complex Langevin with two different initial
conditions, one using purely SU$(3)$ links and the other uses a full
SL$(3,\mathbb{C})$ configuration. The known result, shown as a yellow line in
Figure~\ref{Fig2}, is correctly reproduced. It is worth mentioning that 
in case of dynamic stabilisation using a SL$(3,\mathbb{C})$ start, the unitarity
norm remains small but finite, implying that $B^{c}_{x,\nu} \neq 0$.

\begin{figure}[!ht]
	\centering 
	\begin{minipage}{0.93\linewidth}
	\centering
	\includegraphics[width=\linewidth]{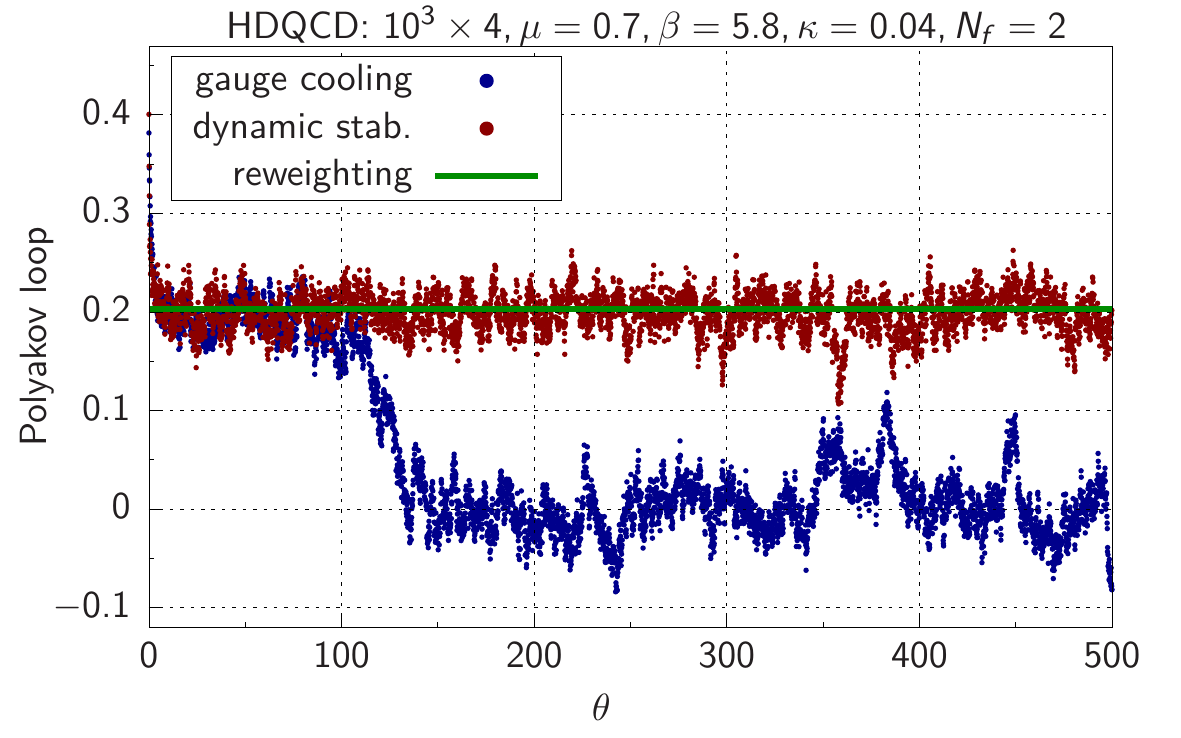}
	\end{minipage}
\caption{Polyakov loop as a function of Langevin time $\theta$. The simulation using 
just gauge cooling (blue) changes to a wrong result at $\theta \sim 100$.
The one including dynamical stabilisation (red) remains at the correct value indicated by
reweighting (green).}
\label{Fig3} 
\end{figure}
For QCD in the limit of heavy quarks (HDQCD), we have seen
examples~\cite{Aarts:2016qrv}, in which complex Langevin simulations diverted away 
after an arbitrary Langevin time from the expected results. The
change of behaviour coincides with the unitarity norm exceeding
$\mathcal{O}(0.5)$.
\begin{figure}[!ht]
	\centering
	\begin{minipage}{0.93\linewidth}
	\centering
	\includegraphics[width=\linewidth]{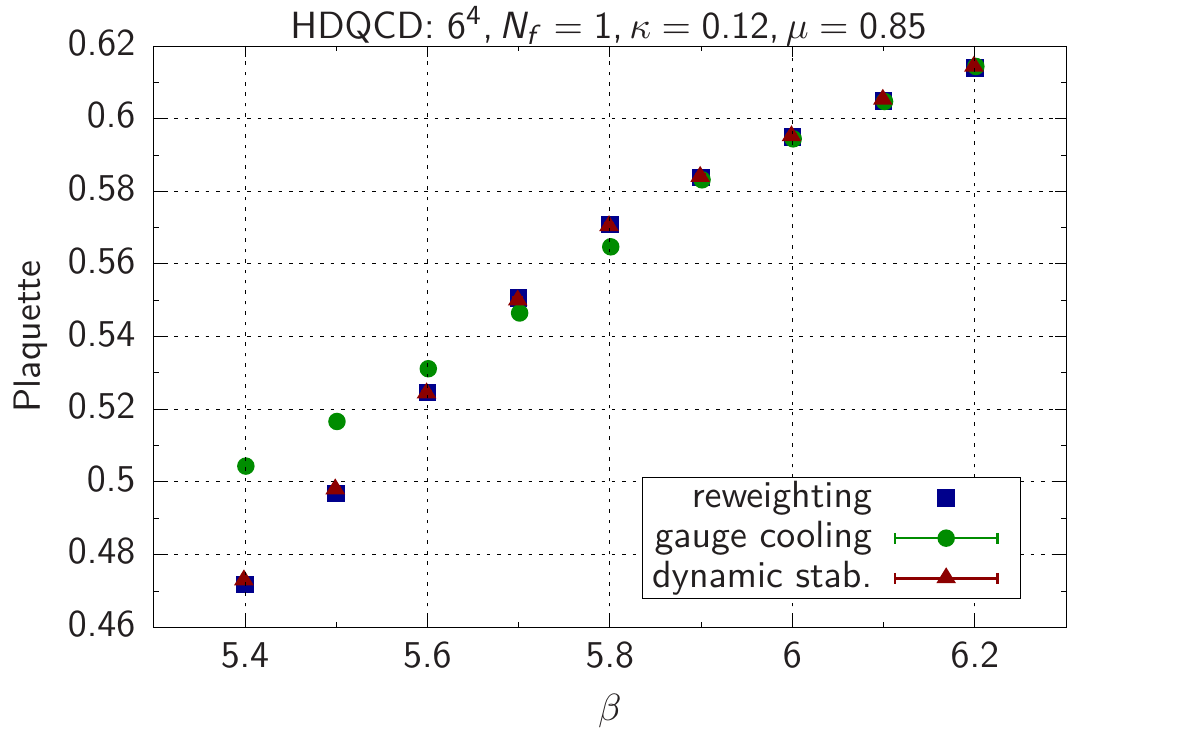}
	\end{minipage}
\caption{The plaquette as a function of the gauge coupling $\beta$. At small 
gauge couplings, gauge cooling, using an adaptive number (up to 20) of
cooling steps, shows significant deviations from reweighting data taken
from~\cite{Seiler:2012wz}. Including dynamic stabilisation improves convergence for all available $\beta$. }
\label{Fig4} 
\end{figure}
An example is shown in Figure~\ref{Fig3} in blue. The simulation parameters
are listed at the top of the plot. After adding the dynamic stabilisation force to
Langevin dynamics we find agreement with reweighting for all Langevin
time. Different gauge couplings are shown in Figure~\ref{Fig4}. 
We find significant deviations between gauge cooling and reweighting results, taken
from~\cite{Seiler:2012wz}, using adaptive gauge cooling over $500$ Langevin time for
$\beta \lesssim 5.8$. Dynamic stabilisation improves the convergence of HDQCD
simulations and enable simulations for small gauge couplings. Further test on
different models and scenarios are subject to future studies.

\section{Conclusions and Outlook}
 
We presented here two tests of our newly proposed method of dynamic
stabilisation for pure gauge simulation and the heavy dense approximation of QCD
(HDQCD). In both cases we have seen clear improvements over gauge cooling and very good agreement
with reweighting. Further tests will validate if this method is suitable for
simulation of dynamical QCD. Here in particular smaller gauge couplings allow
simulations in the confined phase and might enable simulations at low
temperatures. The method, however, remains heuristic and requires adequate
testing and justification. \newline \phantom{stuff}\newline

{\bf Acknowledgements:} 
The authors thank Gert Aarts, D\'{e}nes Sexty, Erhard Seiler and Ion-Olimpiu
Stamatescu for valuable discussions and collaboration. We also appreciate the
reweighting results obtained and provided by D\'{e}nes Sexty and Ion-Olimpiu
Stamatescu. We are grateful for the computing resources made available by HPC
Wales. We acknowledge the STFC grant ST/L000369/1, the Royal Society and the Wolfson Foundation.
FA is grateful for the support through the Brazilian government programme 
“Science without Borders” under scholarship number BEX 9463/13-5.
 

\end{document}